\pdfoutput=1
\documentclass[
reprint,
superscriptaddress,
amsmath,
amssymb,
aps]{revtex4-2}
\usepackage[english]{babel}
\usepackage{amssymb}
\usepackage{hyperref}
\usepackage[letterpaper,top=2cm,bottom=2cm,left=3cm,right=3cm,marginparwidth=1.75cm]{geometry}
\usepackage{amsmath}
\usepackage{xcolor}
\usepackage{dcolumn}
\usepackage{bm}
\usepackage{graphicx} 
\usepackage{mathtools}
\usepackage{textcomp}

\begin{document}
%
\title{ Broadband transparent Huygens' spaceplates}

\author{Francisco J. Díaz-Fernández}
\affiliation{Department of Electronics and Nanoengineering, Aalto University, Maarintie 8,Espoo 02150, Finland}
\affiliation{Nanophotonics Technology Center, Universitat Politècnica de València, Camí de Vera s/n, Valencia 46022, Spain}
\author{Luis Manuel Máñez-Espina}%
\author{Ana Díaz-Rubio}%
\affiliation{Nanophotonics Technology Center, Universitat Politècnica de València, Camí de Vera s/n, Valencia 46022, Spain}%
\author{Viktar Asadchy}
\affiliation{Department of Electronics and Nanoengineering, Aalto University, Maarintie 8,Espoo 02150, Finland}

\date{\today}

\begin{abstract}
Spaceplates have emerged in the context of nonlocal metasurfaces, enabling the compression of optical systems by minimizing the required empty space between their components. In this work, we design and analyze spaceplates that support resonances with opposite symmetries, operating under the so-called Huygens' condition. Using the temporal coupled-mode theory, we demonstrate that the spatial compression provided by Huygens' spaceplates is twice that of conventional single-resonance counterparts. Additionally, they can support broader operational bandwidths and numerical apertures, facilitating the reduction of chromatic aberrations. Moreover, Huygens' spaceplates maintain nearly full transparency over a wide frequency and angular range, allowing their straightforward cascading for multi-frequency broadband operation. Finally, we propose a physical implementation of a Huygens' spaceplate for optical frequencies based on a photonic crystal slab geometry. 
\end{abstract}
\maketitle
\section*{Introduction}
The advent of spaceplates provides new opportunities in the field of non-local metasurfaces~\cite{Reshef2021,Guo2020,Shastri2022,Overvig2022,Kolkowski2023}, allowing to decrease unavoidable free-space regions in an arbitrary optical system (e.g., air gaps between two neighboring lenses). This route of minimizing the overall thickness of optical systems drastically differs from the conventional approaches where the reduction is made in terms of the lens or image sensors thickness~\cite{Pan2022,Miller2023,Chen2016,Lalanne2017,Khorasaninejad2016,Banerji2019,Yu2014,Monticone2013,Khorasaninejad2017,Arbabi2015,Zhou2023}.
For its operation, a spaceplate must have a specific (quadratic) dependence of the complex phase of the transmission coefficient with respect to the incidence angle. At the same time, the magnitude of the transmission coefficient must be close to the unit in a sufficiently broad range of frequencies and incident angles.  

Recently, several material platforms were proposed to realize a spaceplate, including  photonic crystal slabs~\cite{Guo2020,Long2022}, multilayer structures~\cite{Reshef2021,Page2022,Pahlevaninezhad2023,Sorensen2023}, and  Fabry-Pérot cavities~\cite{Chen2021,Mrnka2022,Mrnka2023}.
The limits of operation of these structures have been studied both numerically \cite{Page2022} and analytically \cite{Chen2021, Mrnka2022}, showing an intrinsic trade-off relation between the maximum compression factor of the spaceplate and its numerical aperture (NA). As demonstrated in these works, to overcome the trade-off, one needs to increase the number of coupled resonances in the structure.  
Furthermore, there are stringent limitations on the operational bandwidth of spaceplates providing a given compression ratio~\cite{Shastri2022}. 


Cascading Fabry-Pérot structures have been proposed to increase the number of resonances and surmount the aforementioned limitations~\cite{Chen2021}. These resonances have the same symmetries of the electric type. Although this approach increases the compression ratio, the structure becomes less and less transparent outside the operating range. Moreover, cascading many layers leads to a more complex manufacturing process.
Therefore,  the cascades of Fabry-Pérot structures so far have been implemented only for microwave frequencies~\cite{Mrnka2022,Mrnka2023}. 

In this work,  we propose and design spaceplates supporting resonances of opposite even (electric) and odd (magnetic) symmetries. In particular, by spectrally overlapping the lowest order electric and magnetic dipolar resonances in a photonic crystal slab, we demonstrate the possibility of reaching the so-called Huygens' condition that was previously explored in different contexts~\cite{Decker2015,Epstein2016,Yang2019,Pfeiffer2013}. Due to Huygens' condition, the spaceplate exhibits almost no reflections in a broad frequency range while near the resonance, the phase of transmission varies rapidly with frequency and incidence angle.
Such Huygens' spaceplates exhibit double enhancement of the compression factor compared to the best attainable values for conventional single-resonance spaceplates. Furthermore, we show that Huygens' spaceplates can provide higher operational bandwidths and remain nearly transparent in a broad frequency range.
Using the temporal coupled-mode theory~\cite{Suh2004} (CMT), we propose a realistic optical design of a Huygens' spaceplate. 
It should be noted that a related idea of double-resonance spaceplates was very recently suggested in~\cite{Dugan2023}, however,  the structure there was described by a surface susceptibility model assuming zero structural thickness, and no realistic optical design was proposed. 



\section*{Results}
\subsection*{Huygens' spaceplates}
\begin{figure}[ht]
\centering
\includegraphics[width=0.5\textwidth]{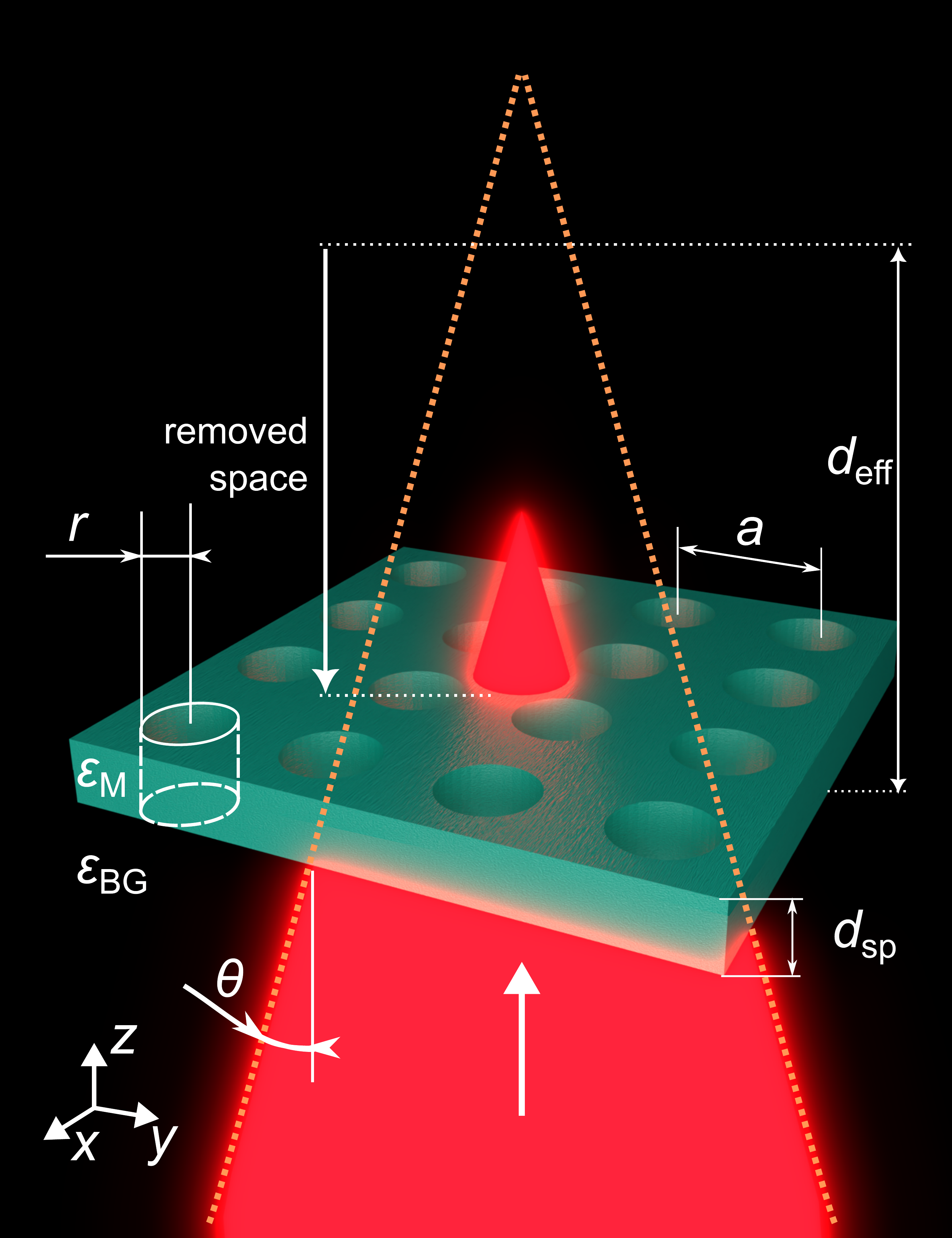}
\caption{\label{fig:SP_concept} Conceptual illustration of a photonic crystal slab operating as a spaceplate. The slab with permittivity $\varepsilon_\text{M}$ and thickness $d_\text{sp}$ comprises a square-lattice array of holes with radius $r$ and lattice period $a$. The geometric size of the holes and spacing between them is not in the scale compared to the optical beam size. 
The spaceplate effectively compresses the air space in an optical system by $d_{\rm eff} - d_{\rm sp}$ thickness. In the case when the spaceplate is illuminated by a converging beam (shown by an orange dotted line), its operation results in the shift of the beam focal point towards the light source. 
The maximum working angle $\theta$  defines its NA. The bottom white arrow indicates the direction of light propagation.}
\end{figure}

The spaceplates allow reducing the size of the optical devices, with the reduction ratio defined as the ratio between the effective distance that the spaceplate emulates and its width itself ($R=d_\text{eff}/d_\text{sp}$, see Fig.~\ref{fig:SP_concept}).
The range of angles at which the spaceplate operates (from 0 to $\theta_\text{max}$ from the normal direction) defines its numerical aperture NA $= \sin\theta_\text{max}$.
To mimic light propagation over some distance in  free space, the spaceplate must have a specific angular dependence of light transmission. On the one hand, different spatial frequencies of light (plane wave components propagating at different angles $\theta$  with respect to the normal) must acquire different phases, according to a quadratic law. On the other hand, the transmission magnitude must be angle-independent and close to the unity (to avoid parasitic reflections). Free-space behavior can be expressed as a transmission function ($t_{\rm fs}$) with dependence on the transversal plane wave components ($\mathbf{k}_{\rm t}$) using the paraxial approximation as follows \cite{Guo2020}:
\begin{align}
     t_{\rm fs}(\omega, \mathbf{k}_{\rm t})\approx  \exp\left[ i \dfrac{\lambda d_\text{eff}}{4 \pi} |\mathbf{k}_{\rm t}|^2 +\phi_0\right],\label{eq1}
\end{align}
being $\phi_0$ the global phase,  $\lambda$ the wavelength,  and $\omega$ the angular frequency. 

\begin{figure*}[]
\centering
\includegraphics[width=1\textwidth]{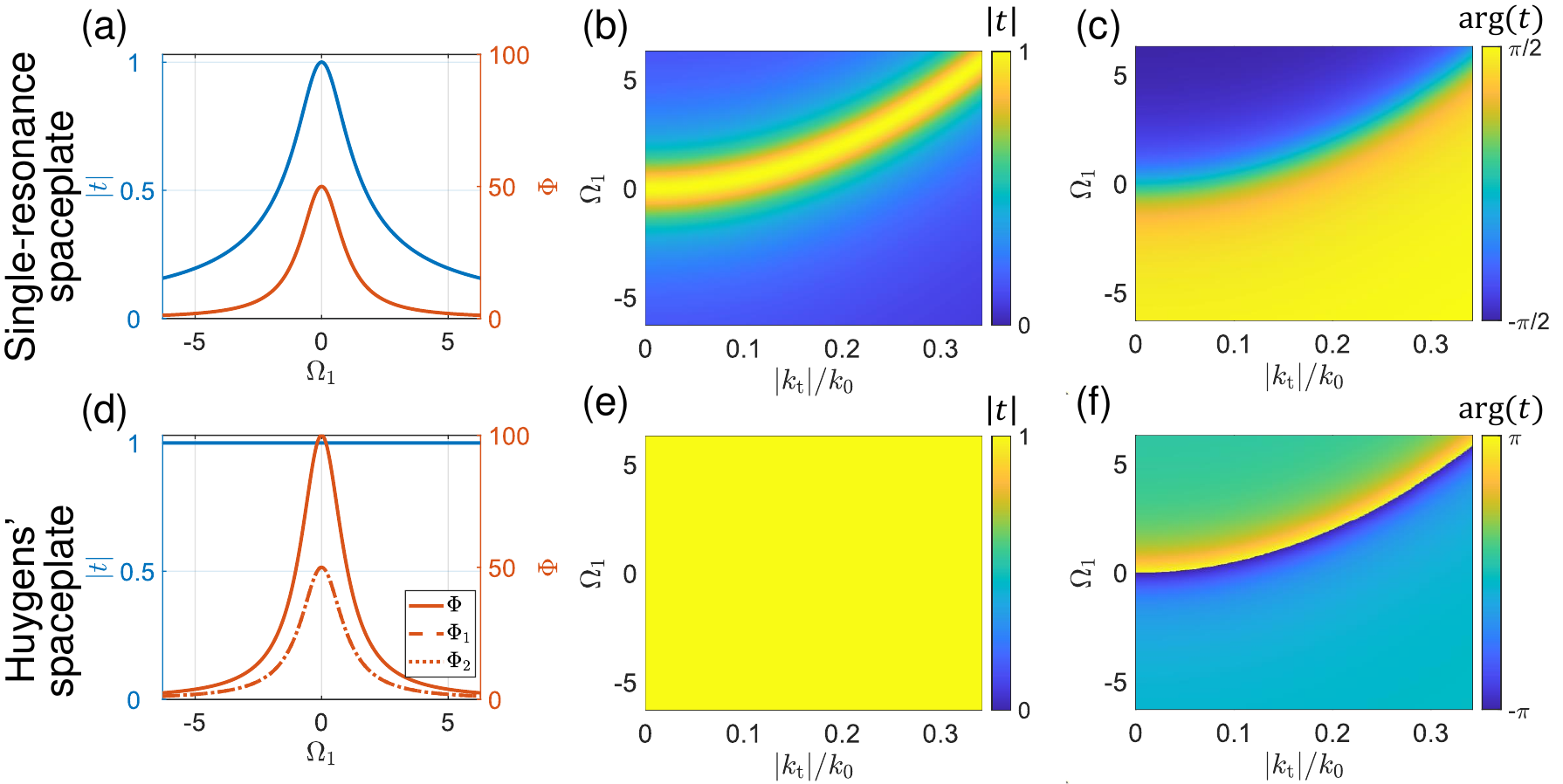}
\caption{\label{fig:best_case_t} 
Comparison of single-resonance and double-resonance (Huygens') spaceplates designed to provide maximum compression parameter $\Phi$ with high transmission coefficient $|t|\approx 1$ at a given frequency.
(a) Transmission coefficient amplitude at normal incidence and compression parameter versus frequency for a spaceplate with single even (electric) resonance. Parameters of the resonance are $\gamma_1= 0.2 \omega_1$,  $\alpha_1=20 \omega_1$, and $t_d=0.01$. (b),(c) Complex transmission coefficient for different frequencies and angles of incidence. (d)--(f) Same for a Huygens' spaceplate with $\omega_2=\omega_1, \gamma_2=\gamma_1, \alpha_2=\alpha_1$, and $t_d=1$. 
}
\end{figure*}
To evaluate to what extent a multi-resonant structure with two resonances mimics the response described by Eq. \ref{eq1}, a coupled-mode model has been developed. Considering the resonances to be orthogonal, the first mode with even symmetry and the second with odd symmetry, the transmission coefficient of a structure can be expressed as follows \cite{Suh2004}: 
\begin{align}
   t_{\rm sp}(\omega, \mathbf{k}_{\rm t})= t_{\rm d} -  \dfrac{\gamma_1(r_{\rm d}+t_{\rm d})}{i[\omega-\omega_1(\mathbf{k}_{\rm t})]+\gamma_1}+\notag\\
    +  \dfrac{\gamma_2(r_{\rm d}-t_{\rm d})}{i[\omega-\omega_2(\mathbf{k_{\rm t}})]+\gamma_2}, \label{eq:cmt_twores}
\end{align}
where $t_{\rm d}$ and $r_{\rm d}$ are the direct transmission and reflection (in the absence of the resonances in the structure), $\omega_{1}(\mathbf{k}_{\rm t})$ and $\omega_{2}(\mathbf{k}_{\rm t})$ are the resonance frequencies for the even and odd modes, respectively, and $\gamma_{1} $ and $\gamma_{2}$ are the widths of these resonances (we assume them being angle-independent). 
Next, we introduce two auxiliary parameters which denote deviations of the operational frequency $\omega$ from a given resonance frequency normalized by the resonance width, that is, $\Omega_{1}(\omega, \mathbf{k}_{\rm t}) =[\omega -\omega_{1}(\mathbf{k}_{\rm t})]/\gamma_{1}$ and $\Omega_{2}(\omega, \mathbf{k}_{\rm t}) =[\omega -\omega_{2}(\mathbf{k}_{\rm t})]/\gamma_{2}$. Defining additionally parameter $q=i r_{\rm d} /t_{\rm d} ( q \in \mathbb{R}$ for lossless slabs), one can write Eq. \eqref{eq:cmt_twores} as
\begin{align}  
t_{\rm sp}(\Omega_1,\Omega_2)= t_d \dfrac{1+\Omega_1(\Omega_2-q)+\Omega_2 q}{(\Omega_1-i)(\Omega_2-i)}. \label{eq:tOmega}
\end{align}

Since the two resonance frequencies might depend differently on the angle of incidence, we can express the frequency dispersion of each resonance for small incidence angles via $\alpha_{1}$  and $\alpha_{2}$ parameters as $\omega_{1}(\mathbf{k}_{\rm t}) \approx \omega_{1}(0) + \alpha_{1} |\mathbf{k}_{\rm t}|^2$ and $\omega_{2}(\mathbf{k}_{\rm t}) \approx \omega_{2}(0) + \alpha_{2} |\mathbf{k}_{\rm t}|^2$. Introducing the Taylor-series expansion (see Sec. 1 of the Supplementary Information), the transmission phase for a structure containing two orthogonal even and odd resonances could be expressed as follows
\begin{align} 
\arg[t_{\rm sp}(\omega,\mathbf{k}_{\rm t})]=\arg[t_{\rm sp}(\Omega_{0,1}, \Omega_{0,2})]  +\notag\\+|\mathbf{k}_{\rm t}|^2\left(\dfrac{\alpha_1}{\gamma_1+\gamma_1\Omega_\text{0,1}^2}+\dfrac{\alpha_2}{\gamma_2+\gamma_2\Omega_\text{0,2}^2}\right),\label{eq:argt}
\end{align}
where $\Omega_{0,1}$ and $\Omega_{0,2}$ represent the frequency detuning factors for the normal incidence as  $\Omega_\text{0,1}  = [\omega-\omega_\text{1}(0)]/ \gamma_1 $
and $\Omega_\text{0,2}  = [\omega-\omega_\text{2}(0)]/ \gamma_2 $.


    Using Eqs. \ref{eq:tOmega} and \ref{eq:argt}, the transmission produced by the spaceplate with two resonances can be written in the form of transmission through free-space slab described by Eq. \eqref{eq1}, that is,
\begin{align}
t_{\rm sp}(\omega,\mathbf{k}_{\rm t})= t_{\rm sp}\left[\Omega_\text{0,1},\Omega_\text{0,2}\right] \exp{ \left[i\Phi |\mathbf{k}_{\rm t}|^2\right]}, \label{eq:finalt}
\end{align}
where  we introduce  compression parameter $\Phi$ as 
\begin{align}
\Phi=   \dfrac{\alpha_1}{\gamma_1+\gamma_1\Omega_\text{0,1}^2}+\dfrac{\alpha_2}{\gamma_2+\gamma_2\Omega_\text{0,2}^2}= \Phi_1 +\Phi_2.\label{eq:phi}\end{align}
From \eqref{eq1} and (\ref{eq:finalt}), one can see that the free-space reduction ratio $R$ is directly proportional to the compression parameter $\Phi$ as $R =4\pi \Phi /(d_\text{sp} \lambda)$. 
From Eq. \eqref{eq:phi}, we can see that the compression parameter for two-resonance spaceplates can be divided into two terms, one term being related to the even resonance and another to the odd resonance. Due to their arithmetic summation, the total compression parameter can become double that for a single-resonance spaceplate~\cite{Dugan2023}.  
This result of doubling of the compression parameter due to combining even and odd resonances in a single geometry is somewhat similar to the doubling of transmission phase span for normal incidence in Huygens' metasurfaces~\cite{Decker2015}. 
Hence, Huygens' spaceplates exhibit a significant advantage with a much higher compression parameter (and reduction ratio) than their single-resonance counterparts. 

To achieve the full transmission through the Huygens' spaceplate $|t_{\rm sp}|=1$ at a given frequency, one needs to make the following condition hold at that frequency, as derived from  Eq.~\eqref{eq:tOmega}:
\begin{align}
    q=\dfrac{\Omega_2-\Omega_1}{1+ \Omega_1 \Omega_2}.\label{eq:qomega}
\end{align}

Let us compare the performance of a single-resonance and Huygens' spaceplates versus the incident angle ($|k_\mathrm{t}|/k_0=\sin \theta$) and the frequency of the incident wave $\omega$.
For a fair comparison, we design both spaceplates  to provide the maximum compression parameter with high transparency ($|t|\approx1$). 
For a spaceplate with a single (even) resonance mode operating at frequency $\omega$,  parameter $\Phi$ is maximized if $\Omega_{0,1}(\omega)=0$, while transmission reaches $|t|=1$ if $\Omega_1(\omega,\textbf{k}_t)=1/q$. The former and latter conditions come from Eqs.~(\ref{eq:phi}) and (\ref{eq:qomega}), respectively, assuming that $|\Omega_2| \rightarrow \infty$ (the second resonance is far apart). 
Therefore, to satisfy both these conditions, one needs to maximize the value of $q$, resulting in  $t_{\rm d} \rightarrow 0 $. 
In Fig.~\ref{fig:best_case_t}(a), we plot the transmission coefficient and compression parameter for
the optimized single-resonance spaceplate with $t_{\rm d} =0.01 $. The angular behavior of the same spaceplate is depicted in Figs.~\ref{fig:best_case_t}(b) and (c). As observed, while the spaceplate is transparent and maintains a high 
$\Phi$ at normal incidence and frequency $\omega=\omega_1$, it forfeits both characteristics outside its narrow operational bandwidth.
It should be noted that previously suggested single-resonance spaceplates in Ref.~\cite{Guo2020} were designed with compression parameters that were far away from the highest possible ones, allowing for a broader transparency frequency range. 
Furthermore, the trade-off  between transmission amplitude and compression parameter in such spaceplates were  studied in~\cite{Shastri2022}.

In sharp contrast to the previous case, Huygens' spaceplates supporting even and odd resonances provide the unique opportunity to have large compression parameter and high transparency in a wide frequency range. In particular, under Huygens' condition of degenerate resonances of opposite symmetries $\omega_1=\omega_2$, we obtain $\left(\Omega_1=\Omega_2\right)$ that makes the right-hand side of Eq.~ (\ref{eq:qomega}) independent of frequency $\omega$. Therefore, the condition of full transparency (\ref{eq:qomega}) can be satisfied ideally at each frequency if we design the slab to have  $q=0$ ($r_{\rm d} \rightarrow 0$).
Therefore, a spaceplate that satisfies  Huygens' condition can achieve the total transmission for every frequency, including those with $\Phi$ maximized, with the double compression parameter provided by a single-resonance spaceplate (assuming the two resonances to have the same spectral characteristics). 
Figures~\ref{fig:best_case_t}(d)--(f) plot the complex transmission versus incident frequency and angle for the ideal Huygens' spaceplate whose two resonances have equal spectral properties, that is,  $\omega_1=\omega_2$, $\alpha_1=\alpha_2$, and $\gamma_1=\gamma_2$. Here, we assumed $t_{\rm d}=1$, which in practice can be obtained, e.g., by adding a uniform slab of specific thickness behind and/or in front of the spaceplate~\cite{Guo2020}.
One can see that this ideal Huygens' spaceplate achieves total transparency for all angles of incidence and frequencies. The $2\pi$-phase span is exactly the double of that in the single-resonance scenario (compare Figs.~\ref{fig:best_case_t}(c) and (f)), accounting for the twofold increase in free-space compression.

In practice, the bandwidth of the transparency region is always limited by multiple imperfections: other higher-order resonances exist in the structure and perturb the Huygens' balance, 
resonances have a mismatch in terms of $\alpha$ or $\gamma$ parameters, $\gamma$, in fact, depends on $\textbf{k}_\text{t}$, and the bands of the modes become non-parabolic for large $\textbf{k}_\text{t}$. These limitations will be discussed in more detail in the section related to the implementation of the spaceplate. 

Interestingly, for both the considered spaceplates in Fig.~\ref{fig:best_case_t}, operational ${\rm NA} = k_{\rm t,max}/k_0$ is the same. For the ideal Huygens' spaceplate, at the degenerate resonance frequency, $k_{\rm t,max} = 2\pi/\Phi = \pi \gamma_1/\alpha_1$ (see  Sec. 1.2 of the Supplementary Information), being the same value as for the single-resonance spaceplates~\cite{Chen2021}. 
Nevertheless, as shown below,  NA can be increased in Huygens' spaceplates under specific small detuning of the two resonances.  

\subsection*{Beyond Huygens' condition}
\begin{figure*}[ht]
\centering
\includegraphics[width=1\textwidth]{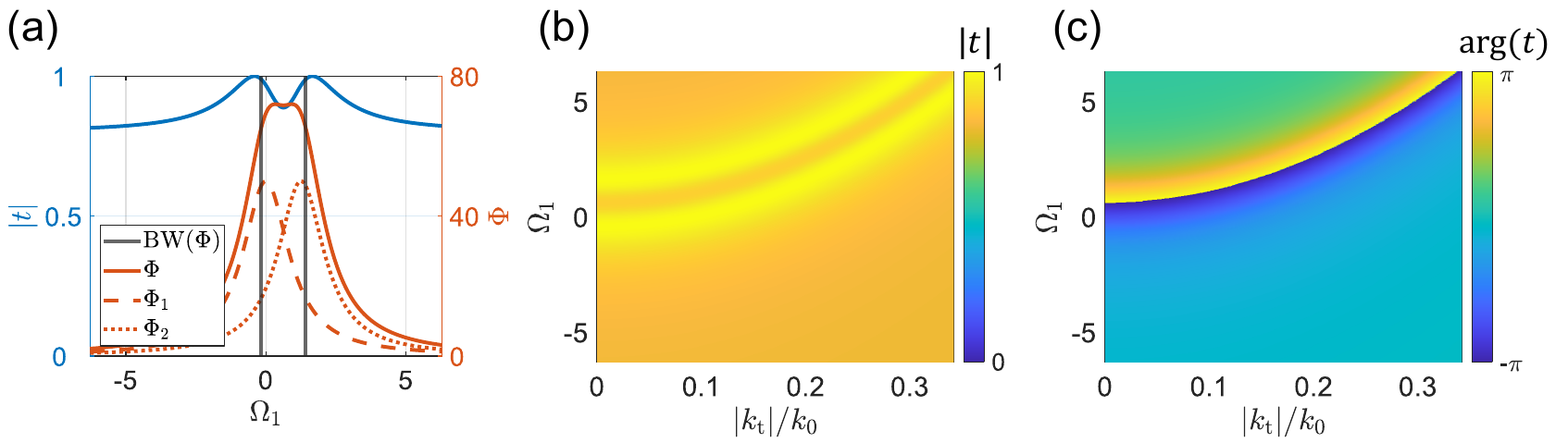}
\caption{\label{fig:chrom} (a) Transmission coefficient at normal incidence (blue) and compression parameter (red) for the spaceplate slightly detuned from the Huygens' condition. The two resonances are separated by $\omega_2-\omega_1 = \Delta\omega=4 \gamma/\pi$, with $\gamma_1=\gamma_2= 0.2 \omega_1$, $t_d=0.8$, and $\alpha_1=\alpha_2=20 \omega_1$. Within the frequency region bounded by the two vertical black lines, the spaceplate has negligible chromatic aberration. 
(b), (c) Complex transmission coefficient for different frequencies and angles of incidence.}
\end{figure*}

Spaceplates, similar to lenses~\cite{Yu2023}, exhibit intrinsic chromatic aberrations, characterized by the phenomenon where light of various frequencies undergoes differing compression parameters $\Phi(\omega)$, as illustrated in  (see Figs.~\ref{fig:best_case_t}(a),(d)). 
These aberrations result in image distortions, which are undesirable in most practical applications. Recent work~\cite{Pahlevaninezhad2023} introduced the theoretical concept of multi-wavelength spaceplates, designed to minimize chromatic aberrations at three specific wavelengths. This was achieved through an elaborate design involving a multilayer structure with several hundred layers. Nevertheless, it worked only for 3 very narrow (nearly \textit{discrete}) frequency regions, remaining highly opaque for light at other frequencies. 

Here, we discuss an alternative approach for overcoming chromatic aberrations in spaceplates by partial overlapping of the resonant modes with opposite symmetries. This approach allows us to decrease the aberrations over a \textit{continuous} frequency range. Using the CMT developed in the previous section, we find that by detuning slightly from the Huygens' condition so that  $\omega_2-\omega_1=\Delta \omega \ll \omega_1$, it is possible to achieve a nearly flat region of $\Phi(\omega)$ with negligible frequency dispersion of compression parameter. 
In Fig.~\ref{fig:chrom}, we plot the complex transmission and compression parameter for a spaceplate operating slightly away from the Huygens' condition when $\Delta \omega=4\gamma/\pi$. The detuning results in the drop of transmission amplitude between the two resonance frequencies but we mitigate this drop by slightly decreasing the background transmission to $t_{\rm d}=0.8$ (other parameters stay the same as in Fig.~\ref{fig:best_case_t}(d)). Although the total transparency is achieved now only at two frequencies, the transmission magnitude is not less than 0.8 for the entire spectrum. 
We define the compression bandwidth  BW($\Phi$) as the frequency range where  $\Phi(\omega) \geq 0.9\Phi_\text{max}$ (black vertical lines in Fig.~\ref{fig:chrom}(a)). For the spaceplate analyzed in
Fig.~\ref{fig:chrom}(a),  BW($\Phi$) is increased compared to the spaceplate in \ref{fig:best_case_t}(d) by around 2.5 times.

\begin{figure}[ht]
\centering
\includegraphics[width=0.49\textwidth]{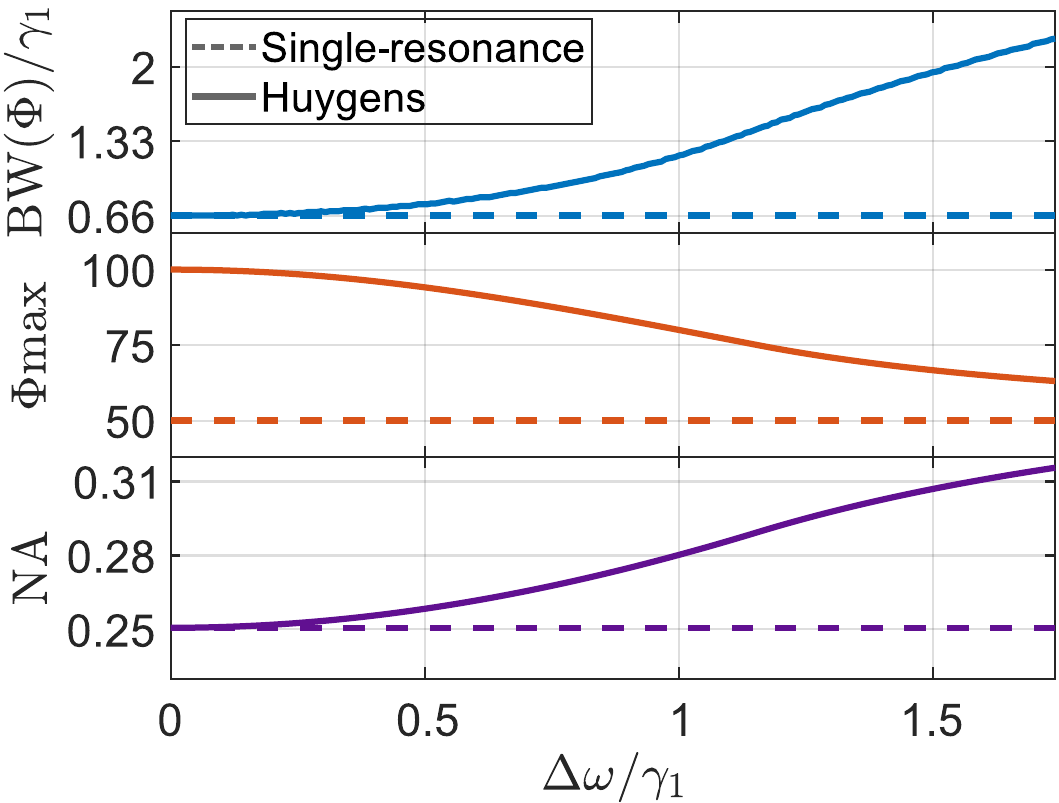}
\caption{\label{fig:Deltaomega} 
Comparison of the chromatic-abberations-free bandwidth BW ($\Phi$), maximum compression parameter $\Phi_{\text{max}}$, and NA of single-resonance (dashed lines) and Huygens' (solid lines) spaceplates. 
The data are plotted versus the frequency separation $\Delta \omega$ for Huygens' spaceplates. The other parameters of the spaceplates are the same as in the caption of Fig.~\ref{fig:chrom}.} 
\end{figure}

Figure~\ref{fig:Deltaomega} depicts a numerical study comparing the BW($\Phi$), $\Phi_\text{max}$ and NA versus the separation of resonances $\Delta \omega$. 
Here two advantages of the double-resonance spaceplates over single-resonance ones can be seen. On the one hand, the achromatic region BW($\Phi$) is increased (blue curves). On the other hand, the purple curve shows how the NA is also increased with the separation of the resonances. It should also be mentioned that the maximum compression parameter becomes reduced when $\Delta \omega \neq 0$. However, it remains higher than that of the single-resonance spaceplate, as illustrated by the red curves. Thus, Huygens' spaceplates provide the simultaneous enhancement of all three parameters by selecting the appropriate distance between their resonances.

\subsection*{Implementation}
 \begin{figure*}[ht]
\centering
\includegraphics[width=1\textwidth]{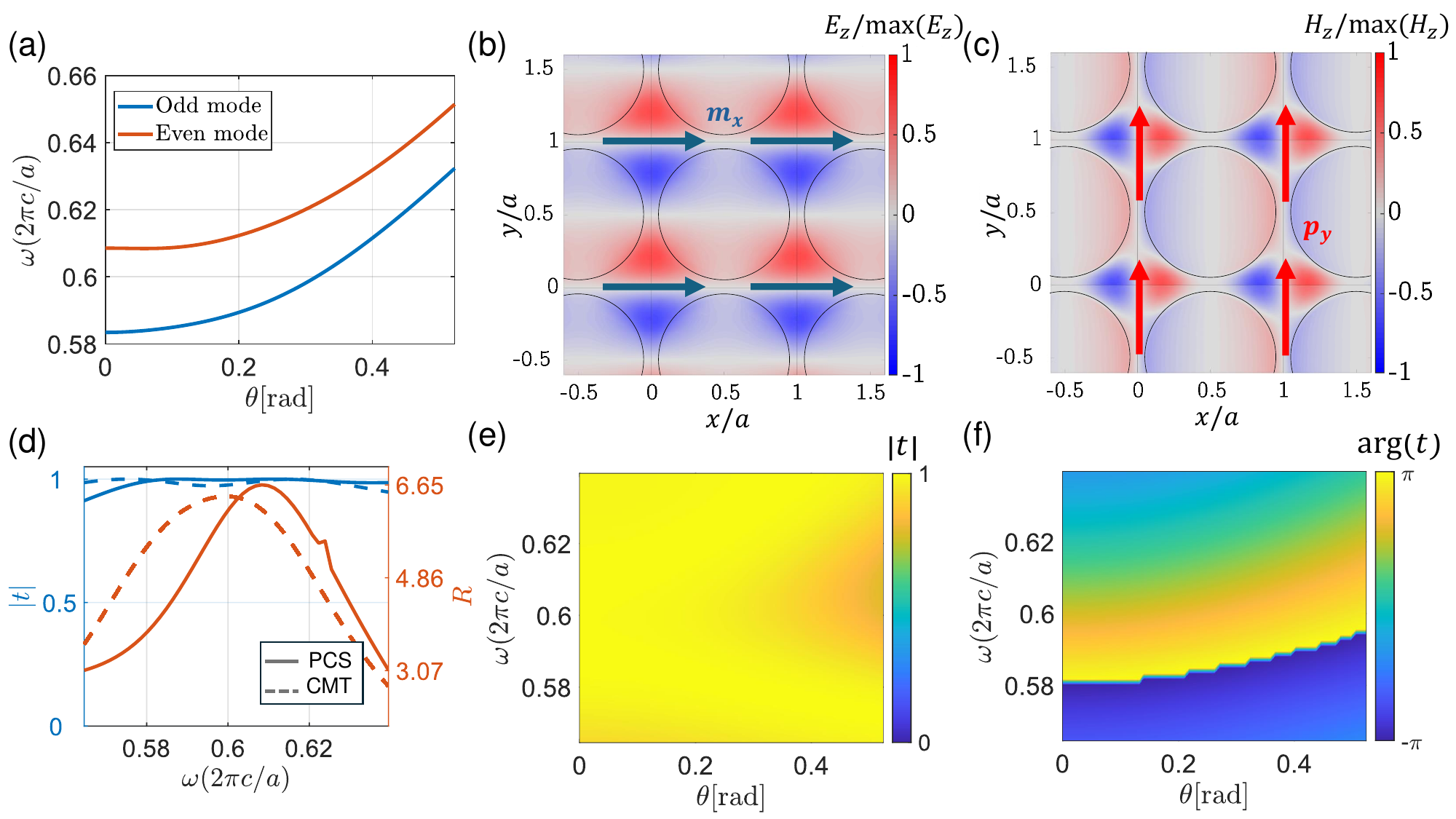}
\caption{\label{fig:Implementation} (a) $\Gamma-X$ band diagram for the studied PCS. (b,c) Eigenmodes (odd and even) for almost overlapped orthogonal resonances: (b) $E_z$ field with the magnetic dipole in blue arrows at $\omega_1$ and (c) $H_z$ field with the electric dipole in red arrows at $\omega_2$. (d) Transmission coefficient for normal incidence (blue) and free space reduction ratio (red) for the two-resonance spaceplate presented via PCS (solid lines) and the CMT approach for the obtained parameters via eigenfrequency study (dashed lines). (e) and (f) correspond to the absolute value and phase for the transmission coefficient at different angles of incidence.}
\end{figure*}

In order to implement the proposed Huygens' spaceplate, we need to choose a photonic geometry that supports two closely spectrally located resonant modes of opposite symmetries. Although one natural choice for the spaceplate geometry would be a metasurface consisting of dielectric cylinders with quasi-Mie resonances~\cite{Decker2015}, we found that the curvature of these resonance bands near $\textbf{k}_{\rm t}=0$ is negative, meaning that $\alpha_1<0$ and $\alpha_2<0$. Such curvature cannot provide the space compression effect, rendering $d_{\rm eff}<0$ in~(\ref{eq1}). 

Therefore, for our spaceplate design, we use a photonic crystal slab of width $d_\text{sp}$  including a square lattice of holes with radius $r$, as shown in Fig.~\ref{fig:SP_concept}. The lattice period is $a$, and the slab permittivity is $\varepsilon_{\rm M}$. In contrast to the single-resonance photonic crystal slabs~\cite{Guo2020,Long2022}, here we utilize the slab supporting electric and magnetic overlapped resonances recently proposed in a different context~\cite{Yang2019}.

As proof of the concept, here we design the spaceplate operating near Huygens' condition similar to that shown in Fig.~\ref{fig:chrom}(a). Such design is more straightforward as it operates at $|t_{\rm d}|<1$.
It should be mentioned that it is also possible to design the spaceplate operating exactly at the Huygens' condition like that in Fig.~\ref{fig:best_case_t}(d). However, in this case, due to the requirement of $|t_{\rm d}| \approx 1$, one needs to engineer carefully the background (non-resonant) transmission coefficient, which can be done by adding two uniform material layers around the photonic crystal slab~\cite{Guo2020}.

Based on the CMT, we optimized the dimensions of the spaceplate as follows:
$d_\text{sp}=\lambda_0/6$, $a=0.527 \lambda_0$, $r=0.452 \lambda_0$. We selected  $\epsilon_\text{BG}=1$ and $\epsilon_\text{M}=11.9$ for the permittivities of background and slab, equivalent to those of air and silicon in the near-infrared regime.  
Such a photonic crystal slab supports two resonance modes with opposite symmetries in the considered frequency range, as shown in Fig.~\ref{fig:Implementation}(a). As mentioned above, the modes are slightly spectrally detuned to have a broadband compression parameter with high broadband transmission as the regime in Fig.~\ref{fig:chrom}(a). Both modes correspond to the same transverse-electric polarization with the electric field along the $y$-axis. The field distributions of the two modes are plotted in Figs.~\ref{fig:chrom}(b) and (c), revealing that the lower-frequency mode $\omega_1$ corresponds to the lattice of magnetic dipoles $m$ along the $x$-direction and the higher-frequency mode $\omega_2$ to the lattice of electric dipoles $p$ along the $y$-direction. The dipoles are excited in the high-permittivity regions between the air voids of the photonic crystal slab. 
Thanks to the almost overlapping even and odd resonances, we achieve a high transmission close to the unity for a wide range of frequencies and angles of incidence, as shown in Figs.~\ref{fig:Implementation}(d)--(f). The transmission amplitude curve has a similar shape to the one in Fig.~\ref{fig:chrom}(a). At the same time, the normalized chromatic-abberation-free bandwidth ${\rm BW}(\Phi)/\gamma_1= 1.01$, which is significantly higher than the maximal bandwidth of single-resonance spaceplates, as seen from Fig.~\ref{fig:Deltaomega}. Moreover, the relative bandwidth reaches $2{\rm BW}(\Phi)/(\omega_1+\omega_2))= 5 \%$, is much higher than that of most of the previously proposed narrowband spaceplates~\cite{Shastri2022}.  
The reduction ratio $R$ of the spaceplate reaches the value of 6.65. It should be noted that this value could be further increased by several orders of magnitude by reaching the Huygens' condition using higher-order (quadrupoles, octupoles, etc) resonant modes in the photonic crystal slab. Indeed, their quality factors are significantly higher than those of the dipolar modes, allowing higher reduction ratios.

Next, we fit the transmission coefficient and reduction ratio spectra to the expressions derived from the coupled-mode model. We chose the resonance frequencies $\omega_1 = 0.5834 \times 2\pi c/a$ and $\omega_2 = 0.6085 \times 2\pi c/a$ from Fig.~\ref{fig:Implementation}(a).
The background transmission was estimated to be $t_d=0.818+0.147i$ from the transmission coefficient for a continuous slab of thickness $d_{\rm sp}$ and permittivity $\varepsilon_{\rm M}$ at the frequency $(\omega_1+\omega_2)/2$. 
Furthermore, we chose $\gamma_1=0.05 \omega_1$, $\gamma_2=0.047 \omega_2$, $\alpha_1 = 0.18 \times c a/ 2\pi$, $\alpha_2 = 0.1856 \times c a/ 2\pi$. The transmission and compression parameter spectra are shown in Fig.~\ref{fig:Implementation}(d) with dashed lines. One can see a noticeable deviation between the spectra from the full-wave simulations and the coupled-mode model. The deviation can be explained by the fact that the considered photonic crystal slab also possesses high-order resonance modes that are not taken into account in the CMT. Moreover, the background transmission coefficient was calculated merely approximately and assumed frequency-independent. Nevertheless, the CMT provides a good qualitative description of the considered process.


According to the simulated results in Fig.~\ref{fig:Implementation}(d), the highest reduction ratio of $R=6.65$ is achieved at $\omega = 0.6094 \times 2\pi c/a$. For this frequency, the NA reaches as high as 0.32 ($\theta_\text{max} = 18.5^{\circ}$) (see Sec. 3 of the Supplemental Information). Such a high value of NA became possible due to the slightly shifting from Huygens' condition (see Fig.~\ref{fig:Deltaomega}). Moreover, for the same spaceplate geometry but at a frequency of $\omega = 0.6262 \times 2\pi c/a$, 
we achieve  ${\rm NA}=0.54$ $ (\theta_\text{max}=33^{\circ}$) with slightly reduced $R=4.89$ (Fig.~S1 in the Supplemental Information). This result is far better than that in known single-resonance spaceplates with similar values of $R$, including those based on multilayer structures~\cite{Reshef2021} (NA = 0.29) or Fabry-Perot resonators~\cite{Chen2021} (NA = 0.33). Higher NA are important for many realistic applications, such as cameras in smartphones or professional imaging equipment~\cite{DelMastro2022}.

It is important to make a comparison of the presented spaceplate with the previous structures based on the performance factor derived in~\cite{Chen2021}. This factor, expressed as  PF $=(\textrm{NA})^2 d_\textrm{eff}/\lambda$, has a theoretical bound, rendering a trade-off between achievable NA and the reduction ratio. In particular, for single-resonance spaceplates it is limited by PF $< 1$ \cite{Chen2021}, while for the double-resonance scenario, PF $\leq 2$ \cite{Dugan2023}. In our case, due to the small width of the spaceplate $d_{\rm sp}$, ${\rm PF }= 0.11$ for the frequency $\omega = 0.6094 \times 2\pi c/a$ where $R$ is maximized. At the frequency $\omega = 0.6262 \times 2\pi c/a$ where the NA is maximized, the factor reaches  ${\rm  PF} = 0.24$. On the one hand, these values are much higher than those obtained in single-resonance spaceplates based on photonic crystal slabs, where PF$< 0.01$ \cite{Guo2020}. On the other hand, these factors are below the values obtained in Fabry-Perot spaceplates (PF $\approx$ 0.5 \cite{Chen2021}). However, in contrast to them, our spaceplate supports broadband transparency both in frequency and angles of incidence, which, combined with the high NA, makes our design more optimal for applications where larger bandwidths are preferable.


\section*{Discussion}

Using the CMT, we have demonstrated that Huygens' spaceplates supporting two resonance modes with opposite symmetries provide important significant enhancement of several characteristics compared to single-resonance counterparts: operational bandwidth, reduction ratio, and NA. The first proof-of-principle design confirms the predicted performance. It was shown that departing from Huygens' condition for the two resonances in the spaceplate provides additional advantages, such as the possibility of reducing chromatic aberrations. 


Another important advantage of Huygens' spaceplates is their transparency in a wide range of angles and frequencies. In this aspect, they are similar to Huygens' metasurfaces~\cite{Shaham2023,Radi2015}.
This feature allows one to straightforwardly cascade several different Huygens' spaceplates one after another. Through this cascading, it would be possible to either further extend the operational range of the spaceplate or even to create multi-band spaceplates.  


It is worth mentioning that the results obtained in this paper can be extended to any frequency, especially in the optical domain. To this aim, we proposed an example of a silicon-based photonic crystal slab that can operate as a spaceplate in the near-infrared region. 
The proposed structure is monolayer, leading to the simplicity of its fabrication. It could be fabricated in integrated CMOS-compatible systems. The high NA and straightforward integration possibilities pave the way for the proposed spaceplates to be used in real-world applications, from reducing invasive medical devices to cameras in professional environments or smartphones.

\section*{Methods}
\subsection*{Analytical development}
Along with this work, the analytical development of the CMT has been performed (see Sec. 1 of the Supplementary Information for the complete development). The authors have used Wolfram Mathematica to support the performed development. The numerical results presented in this work have been calculated and presented via Matlab software. 

\subsection*{Full-wave simulations}
The eigenmode analysis was performed in COMSOL Multiphysics. Periodical boundary conditions were applied on the in-plane faces of the structure in conjunction with the Floquet condition along the $x$-axis to obtain the $\Gamma - \rm X$ band diagram. Scattering boundary conditions were applied in out-of-plane faces. The system is, therefore, open, and eigenfrequencies have non-zero imaginary parts, even in the absence of material losses.

To study the behavior of the proposed spaceplate (transmission coefficient), the numerical simulations of the photonic crystal slabs were performed using the electromagnetic wave frequency domain (emw) solver in COMSOL Multiphysics. 
Two periodic boundary conditions were assigned, one in the $x$-direction and another along the $y$-direction, with the Floquet ports defined as normal to $z-$axis. 
The angle of incidence $\theta$ has been swept in the $xz$-plane, with the electric field polarized along the $y$-direction. The diffraction orders were evanescent for the entire range of considered incident angles. 
We used Matlab to fit the phase behavior of the photonic crystal slab with the empty space of an effective distance $d_\text{eff}= R d_\text{SP}$ with the tolerance of 2\% (see Sec. 3 of the Supplementary Information). 
\\
\section*{Data availability}
Data used in this study are available from the corresponding author upon request.

\section*{Acknowledgements}
The authors would like to thank Prof. Sergei Tretyakov, Prof. Constantin Simovski, and Dr. Francisco S. Cuesta for the fruitful discussions.
F.J.D.-F. acknowledges the Next Generation EU program, Spanish National Research Council (Ayuda Margarita Salas), and Universitat Politècnica de València (PAID-06-23). L.M.M-E. acknowledges Universitat Politècnica de València (PAID-01-23). A.D.-R. acknowledges the Beatriz Galindo excellence grant (grant No. BG-00024) and Generalitat Valenciana PROMETEO Program (CIPROM/2022/14).  V.A. acknowledges the Academy of Finland (Project No. 356797) and  Research Council of Finland Flagship Programme, Photonics Research and Innovation (PREIN), decision number 346529, Aalto University. 

\section*{Authors Contributions}
A.D.-R. and V.A. conceived the idea. F.J.D.-F., L.M.M-E., and V.A. carried out the theoretical development. F.J.D-F. performed the numerical results, designed and carried out the simulations of the implemented design. L.M.M-E. performed the eigenvalues study. F.J.D-F., A.D-R., and V.A. co-wrote the paper. V.A. supervised the project.
\bibliographystyle{naturemag}
\bibliography{Spaceplates}

\end{document}